# EFFECTS OF MINDFULNESS ON PERCEIVED STRESS LEVELS AND HEART RATE VARIABILITY

**Gwyneth Wesley Rolph**

## ABSTRACT

Mindfulness has become increasingly popular as a method for building resilience against stress in both clinical and healthy populations.  This study sought to investigate the effects of mindfulness training on perceived levels of stress and heart rate variability in students.  Thirteen Westminster University undergraduates enrolled in the Clinical Applications of Mindfulness module undertook an eight-week mindfulness training course and their stress levels were measured before and after the course using the Perceived Stress Scale and emWave2® heart rate variability biofeedback device.  A case study also examined one participant's data in further depth.  In contrast to earlier research, there was a slight increase in perceived stress scores after training.  There was also an unexpected slight reduction in heart rate variability scores.  However, the results were not statistically significant and no definitive conclusions can be inferred from this study about the efficacy or otherwise of mindfulness as an intervention for stress.  Nevertheless, mindfulness remains an interesting line of enquiry, and suggestions for future research are proposed.

## INTRODUCTION

### *Mindfulness and its applications*

Since mindfulness was introduced to the Western world as a secular alternative to the Buddhist traditions and meditative practices at its roots, its popularity has increased dramatically over the last two decades.  One popular definition of mindfulness is "*the awareness that emerges through paying attention on purpose, in the present moment, and non-judgementally to things as they are*" (Williams, Teasdale, Segal & Kabat-Zinn, 2007).

Mindfulness is a state, not a set of exercises or character trait; however the practice of certain exercises can induce a mindful state (Davis & Hayes, 2011).  Kabat-Zinn (1990) introduced the medical world to mindfulness as a new healing paradigm where patients were encouraged to explore the role of their own minds in illness and health.  Mindfulness practice has since then been applied extensively to treat a wide variety of clinical conditions such as depression, anxiety, eating disorders, personality disorders, addictions, chronic pain, and PTSD (Baer, 2006).  Such mindfulness-based exercises are now extensively employed in therapies such as acceptance and commitment therapy (Hayes, Strosahl & Wilson, 1999), Dialectic Behaviour Therapy (Linehan, Schmidt, Dimeff, Craft, Kanter & Comtois, 1999), mindfulness-based stress reduction (MBSR) (Kabat-Zinn, 1990) and mindfulness-based cognitive therapy (MBCT) (Teasdale, Williams & Segal, 2014).  Such programmes have also been increasingly used in non-clinical populations for overcoming stress and enhancing day-to-day resilience (Kabat-Zinn, 1990).  It is this latter application of mindfulness training that is of particular interest in this study.

*Mechanisms of stress*

Selye (1955) defined stress in terms of broadly generalised responses to *stressors* – internal or external pressures or demands on an organism, and the word *stress* has now entered popular parlance to refer to any mental or emotional reaction to adverse circumstances. When an organism perceives a potential threat, the sympathetic branch of the autonomic nervous system is activated, suppressing the activity of the parasympathetic branch and initiating the sympathomedullary (SAM) pathway (Bear, Connors & Paradiso, 2007). Signals from the hypothalamus activate the adrenal medulla, releasing adrenaline to prepare the organism for immediate fight or flight. Once the threatening situation has passed, parasympathetic activity brings the body's systems back to homeostasis. However, human behaviour and thought patterns can often induce reactive states of chronic stress in which individuals can become stuck (Kabat-Zinn, 1990). This long-term stress activates the hypothalamic pituitary adrenal (HPA) axis, where signals from the hypothalamus stimulate the pituitary gland to secrete adrenocorticotropic hormone (ACTH), which signals the adrenal glands to produce cortisol (Bear et al., 2007). While the purpose of cortisol is to release a steady supply of blood glucose from the liver to cope with an ongoing stressor, in the longer term this stress response can lead to a broad range of physical complaints such as high blood pressure, arrhythmias, sleep disturbances, digestive disorders, muscular pain, and chronic hyperarousal, but also maladaptive coping mechanisms such as overwork, hyperactivity, overeating, and substance dependency (Kabat-Zinn, 1990).

*Towards a new treatment model*

The answer of traditional medicine to the physical and mental effects of such maladaptive arousal was to treat each sign or symptom of illness or distress individually. Engel (1977) highlighted the need to take into account social and psychological factors as well as medical ones, and treat the patient holistically. The traditional medical model did not explain why, given similar environmental conditions, stressors impacted some people more than others (Kabat-Zinn, 1990). Resilience can be regarded as the ability to cope with stress (Connor & Davidson, 2003), and can be enhanced through various interventions. An important goal of the various mindfulness-based (or mindfulness informed) so-called "third wave therapies" is to teach patients to self-regulate, enhancing their capacity for resilience (Kabat-Zinn, 1990). This has important ramifications for an individual's performance in the workplace or in an educational setting. Yerkes and Dodson (1908) were among the earliest to identify the existence of an optimum level of physiological arousal for optimal performance. Performance can be visualised as an inverted U-shaped curve, suffering due to disengagement and boredom where there is too little arousal, and from over anxiety where there is too much. This has obvious implications for populations such as university students and professionals in high-pressure roles. Institute of HeartMath (2014) describes a model whereby mindfulness training and biofeedback therapy widen the performance curve, so that the individual can perform optimally in a wider range of arousal and enjoy greater resilience to stress.

*Literature review*

A review of the literature on mindfulness for stress reduction revealed a number of studies on nonclinical samples using either the MBSR or MBCT programme. An extensive review of studies comparing the effect of MBSR training on health, life quality, and social functioning in clinical and healthy populations (de Vibe, Bjørndal, Tipton, Hammerstrøm & Kowalski, 2012) found consistent improvements in stress levels and coping strategies in all participating groups. More recently, a meta-analysis of 29 MBSR and MBCT studies to evaluate their effectiveness in healthy participants (Khoury, Sharma, Rush & Fournier, 2015) showed considerable stress reduction and improved quality of life, as well as moderate but consistently lower depression, anxiety and distress scores on a variety of clinical instruments.

Literature on studies in university students included a randomised controlled trial using MBCT as a self-help intervention (Taylor, Strauss, Cavanagh & Jones, 2014), which revealed significant reductions in stress symptoms as well as depression and anxiety. The authors note that students tend towards higher levels of stress than other nonclinical populations; however, interestingly, a ten-week follow-up also revealed that improvements following the training had been maintained. A brief study in university students following a five-week MBSR course (Bergen-Cico, Possemato & Cheon, 2013) also showed improvements on several measures of psychological health despite the brevity of the training.

One criticism of previous research is that a clear mechanism of action of mindfulness training has not yet been established. Gu, Strauss, Bond and Cavanagh (2015) identified a number of hypotheses in their review and meta-analysis. Theoretical mechanisms proposed follow the general themes of increased adaptive coping skills, acceptance, attention regulation, emotional regulation, self-awareness, and stability. While these serve as useful starting points for measuring outcomes of MBSR/MBCT studies, it is unclear whether such interventions directly affect those outcomes or whether other variables are involved.

The difficulty in establishing a clear mechanism underpinning the changes associated with mindfulness practice is not limited to purely psychological variables. Mindfulness practice also produces changes in the central nervous system and autonomic nervous system, although based on research to date no direct causal link can be inferred. Khazan (2016) identified numerous structural changes shown by fMRI scans which are linked to mindfulness practice and known to improve cognitive performance, emotional and behavioural regulation, and social functioning. These include enlargement of the hippocampus, lateral cerebellum, temporoparietal junction and posterior cingulate cortex, increased activation of the anterior cingulate cortex and right insula, and a decrease in size of the right amygdala. Nijjar, Puppala, Dickinson, Duval, Duprez, Kreitzer and Benditt (2014) reported improved cardiac sympathovagal tone during MBSR training, as measured by heart rate variability measurements. The study sought to determine whether such changes could be observed during meditation, compared to the use of controlled respiration alone. While meditators did show increased autonomic balance, the authors acknowledged the difficulties inherent in isolating the effects of mindfulness meditation practice from the effects of breath control, which is known to increase heart rate variability. Further clarification of the precise mechanisms is required.

### *The current study*

Since stress, as discussed above, comprises both psychological and physiological components, the rationale for this current study was to investigate the effects of mindfulness training on both aspects in a healthy population, and to extend previous research by using both types of measurement in the same study.

The Perceived Stress Scale (PSS) (Cohen, Kamarck & Mermelstein, 1983) was identified as a valid and reliable measure of perceived stress, which was brief (10 items) and straightforward to administer and score.

There were many available physiological measurements associated with stress, but heart rate variability (HRV) was chosen for its simplicity of measurement using a portable biofeedback device. HRV is defined as "...*a measure of neurocardiac function that reflects heart-brain interactions and ANS dynamics*" (McCraty & Shaffer, 2015). The better the heartbeat patterns, breathing patterns, and blood pressure rhythms synchronise, the more *coherent* they are deemed to be (McCraty, Atkinson, Tomasino & Bradley, 2009). A lack of coherence (synchrony) indicates a stress response. Coherence is known to correlate with increased resilience, improved positive affect, increased production of DHEA and reduced cortisol levels, and indicates harmony and balance between the two divisions of the autonomic nervous system.

This study therefore seeks to use the PSS and heart rate variability to measure stress levels in university students before and after a course of mindfulness training. Following the findings of the earlier literature, it is hypothesised that after the mindfulness training (1) Perceived Stress Scale scores will decrease and (2) heart rate variability scores will increase.

# METHOD

## *Design*

This study used a repeated measures experimental design and case study to investigate the effect of an eight week mindfulness-based cognitive therapy course on perceived stress levels and heart rate variability. Perceived stress levels were measured via a self-report questionnaire consisting of rating scales for each question, with some items reversed to control for possible acquiescence bias. A physiological measurement (heart rate variability) was also measured using a proprietary electronic biofeedback device.

## *Participants*

An opportunity sample (*N*=13) was recruited from Westminster University students enrolled in the 5PSYC006W Clinical Applications of Mindfulness module. Nearly 70% of the sample consisted of females (F=9; M=4), which was fairly representative of the module enrollees in general. The participants ranged in age from 19-47 with a median age of 23. Participants were all second-year undergraduates with at least some limited experience of prior research participation. Nine participants completed the study.

## *Materials*

The study used the Perceived Stress Scale (PSS) (Cohen, 1994), a 10-item questionnaire (see **Appendix C**). Respondents rated how often they thought or felt a certain way during the last month using a Likert-type scale where 0=never, 1=almost never, 2=sometimes, 3=fairly often, and 4=very often. Items 1, 2, 3, 6, 9 and 10 asked about feelings of unpredictability, uncontrollability, and overload, and current levels of experienced stress, with low scores indicating low perceived stress and high scores indicating high perceived stress. Items 4, 5, 7 and 8 asked about respondents' confidence in handling life stressors. These items are scored in reverse, so a response of 0 is scored 4, and so on. Possible scores on the PSS range from 0-40, where 0 would be very low and 40 extremely high.

In addition, the study used the emWave2, a pocket-sized electronic device, to measure heart rate variability (HRV). An infrared ear clip detects signals through the earlobe and physiological recordings are saved on the device until they can be uploaded to a computer for analysis using the Emwave Pro software. The Kubios HRV software package was also used to further analyse the data.

**Appendix A** shows the information sheets which were provided to participants on the purpose of the study and on the emWave2 device.

### Procedure

All participants ($N$=13) completed the Perceived Stress Scale questionnaire, and two minutes' heart rate variability data were recorded using the emWave2. For this study, the emWave2 was not used as a biofeedback tool, the purpose for which the device is marketed, but purely as a measurement instrument to sample heart rhythm activity. Participants then followed an eight week long Mindfulness Based Cognitive Therapy course based on the programme devised by Teasdale, Williams and Segal (2014). Each session, which was led by a qualified instructor, lasted approximately 1½ hours and took place once a week. In addition, participants were given "homework" exercises from the Teasdale et al. course manual and accompanying audio guided meditation instructions (2014) to practise during the week and encouraged to keep a reflective log. At the end of the eight week course, the PSS questionnaire was administered to participants for a second time, and a further two minutes' data were recorded using the emWave2.

The questionnaires were scored in accordance with the published instructions, and physiological data recordings from the emWave2 were uploaded to a PC where they could be read with the emWave Pro software. Both data sets were then statistically analysed to compare scores before and after mindfulness training. The physiological measurement variable used for this experiment was the emWave mean coherence score, an arbitrarily defined scale ranging from 0-16 measuring heart rhythm patterns corresponding with ANS activity, where low scores indicate a stress response. The recordings were also exported to the Kubios HRV software package in order to access data analysis features unavailable in emWave Pro. These data were only explored in depth on one participant, however, in the form of a case study.

### Ethical Issues

All participants were informed prior to the course that mindfulness can increase emotional awareness and possible emotional discomfort may arise. Therefore all participants were screened prior to the study for recent mental health issues, bereavement or trauma, physical illness or difficulties, or medication use that may render them particularly vulnerable to these effects. Participants were instructed to inform the mindfulness teacher in confidence of any concerns, and request assistance from appropriately qualified professionals if they required additional support. Participants were informed that data from the mindfulness and perceived stress study would be used by the research team but kept confidential. Participation in the study was voluntary and participants were informed that they could withdraw from the study at any stage.

# RESULTS

***Experimental Data***

The mean Perceived Stress Scale (PSS) score before the mindfulness course was 15.78 with a standard deviation of 5.89.  After the mindfulness training, the mean PSS score was interestingly slightly higher at 16.11, and had a standard deviation of 8.84, which revealed that the distribution of scores was more varied (see **Table 3**, **Appendix D**).  **Figure 1** shows a bar chart illustrating the mean PSS scores before and after mindfulness training.

The skewness and kurtosis values indicated normal distribution of data, so despite the very low number of participants ($N$=13), a paired $t$-test was used because in view of the foregoing it was considered sufficiently robust to analyse the results.  The paired $t$-test showed that the difference in scores before and after training was not statistically significant ($t$=-0.17, $df$=8, $p$=0.43, one-tailed).  The effect size ($d$=-0.04) was very small, although the direction of effect was actually in the opposite direction to what had been hypothesised, increasing instead of decreasing.  The experimental hypothesis that mindfulness training would reduce PSS scores was therefore rejected.

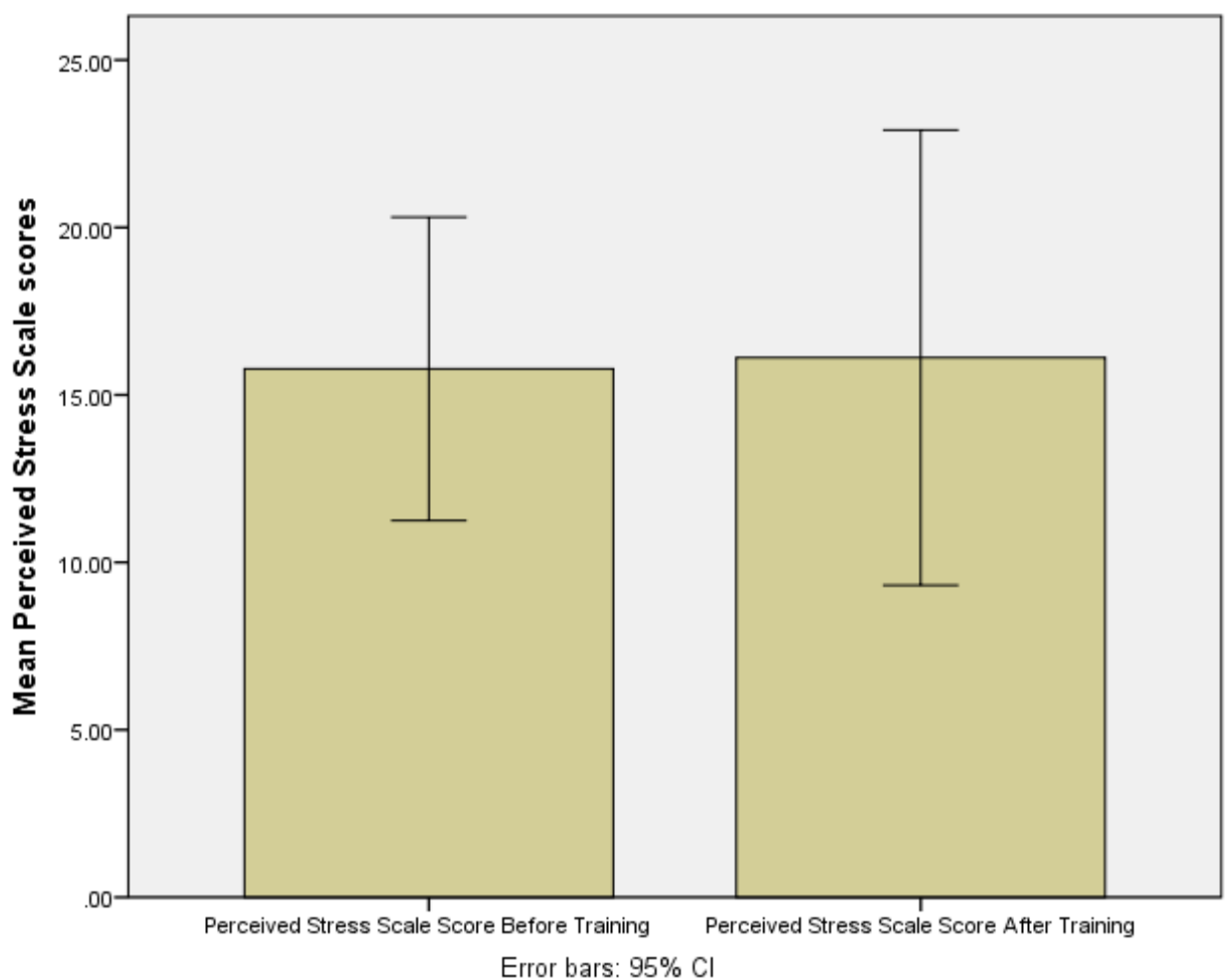

Error bars: 95% CI

**Figure 1: Perceived Stress Scale questionnaire scores before and after mindfulness training**

The mean heart rate variability coherence score before the mindfulness training was 1.01 with a standard deviation of 0.71. Upon completion of the course, the mean heart rate variability coherence score was 0.95 with a standard deviation of 0.89, which was slightly lower than before the training, but the scores were also more varied. (See **Table 5**, **Appendix D**.) **Figure 2** shows a bar chart illustrating the mean coherence scores before and after mindfulness training.

The skewness and kurtosis values showed a non-normal distribution (see **Table 5**, **Appendix D**). A Wilcoxon Signed Ranks test showed that the difference in coherence scores before and after training was not statistically significant ($Z$=-.28, $N$-Ties=8, $p$=0.39, one-tailed) (see **Table 6**, **Appendix D**). Again, similarly to the PSS scores, the HRV coherence scores did not go in the predicted direction, in this case decreasing slightly instead of increasing. The experimental hypothesis that mindfulness training would increase HRV coherence scores was therefore rejected.

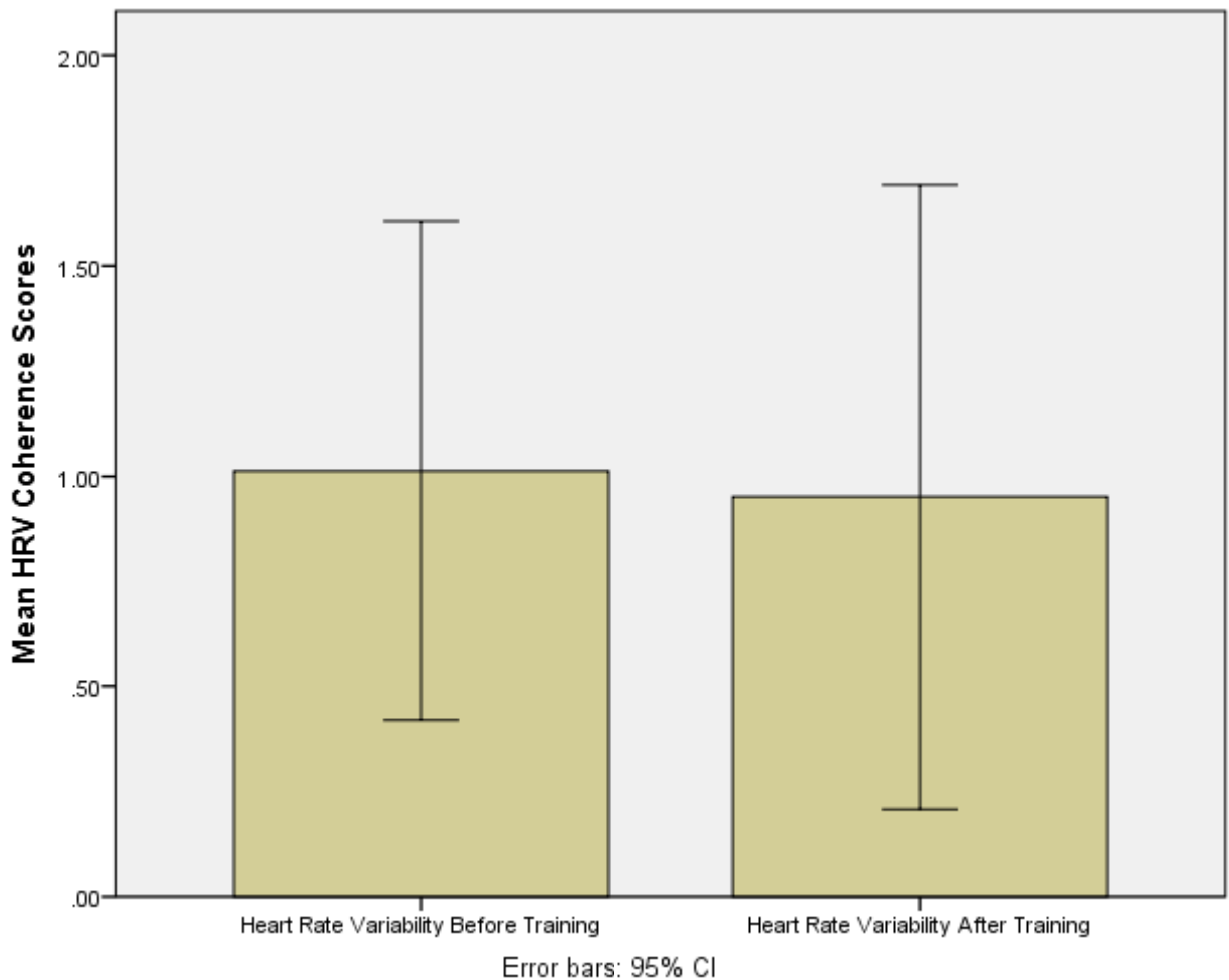

**Figure 2: Mean HRV Coherence scores before and after mindfulness training**

### Case Study

Participant No. 10, a 24-year old female participant, was selected for this case study because the data suggested this was one of the most representative participants in the sample, both demographically and in terms of scores on both variables, with no missing values. This participant's Z-scores on the experimental variables were all within approximately half a standard score from the mean, and these are summarised in **Table 1** below.

**Table 1: Perceived Stress Scale (PSS) and HRV Coherence for Participant No. 10 before and after mindfulness training**

| Experimental Variable | Raw Score | Z-Score | Percentile Rank |
|---|---|---|---|
| Perceived Stress Scale score before training | 18 | 0.39 | 65.17% |
| Perceived Stress Scale score after training | 17 | 0.10 | 53.98% |
| HRV Coherence score before training | 1.1 | 0.35 | 63.68% |
| HRV Coherence score after training | 0.5 | -0.51 | 30.50% |

Further investigation of the recorded physiological data using Kubios HRV software also revealed Participant No. 10's mean heart rate, measured in beats per minute (bpm) and variability measured as the R-R interval (where R is the peak of the QRS complex of the electrocardiogram wave). This participant's heart rate was reduced by over 10 bpm after mindfulness training and had also become more varied. (See **Table 2**.) Although this seemingly contradicts the reduction in coherence score, different processes are being measured (see Discussion) and R-R interval and heart rate variability, while related, are not the same.

**Table 2: Mean Heart Rate and R-R Interval for Participant No. 10 before and after mindfulness training**

| Type of Measurement | Rate |
|---|---|
| Mean Heart Rate before training | 81.48 bpm |
| Mean Heart Rate after training | 70.49 bpm |
| R-R Interval before training | 738.27 ms |
| R-R Interval after training | 855.51 ms |

## DISCUSSION

### Experimental Data

This study hypothesised that after an eight-week Mindfulness-Based Cognitive Therapy course, participants' Perceived Stress Scale (PSS) scores would decrease and their HRV coherence scores would increase. There was in fact a slight increase in PSS scores after the mindfulness training course, while HRV coherence scores similarly contradicted the experimental hypothesis by showing a slight decrease after mindfulness training. These findings do not accord with previous literature on the effects of mindfulness training, which consistently reported reduced stress levels after MBSR or MBCT training (de Vibe et al., 2012; Khoury et al., 2015; Taylor et al., 2014; Bergen-Cico et al., 2013). On the two measures recorded by this study, previous literature suggested that psychological appraisal of stress (Cohen et al., 1983) would have decreased after the mindfulness course and measurement of autonomic nervous system activity (Nijjar et al., 2014) would have shown greater parasympathetic activation, reflected in higher HRV coherence scores.

The findings of the current study would suggest that, contrary to previous research, participants became more stressed after the mindfulness course. Nothing was identified from the literature that would corroborate or explain such an outcome; therefore it is concluded that a combination of confounding variables and a very small sample size most likely contributed to these anomalous results.

Limitations of the study include situational irrelevant variables, participant irrelevant variables and a lack of other possibly relevant information that was not collected as part of the study. Administration of testing both before and after the mindfulness course was performed under time constraints, which may have been distracting or stressful for some participants. Some participants had just written a highly time-pressured examination the previous afternoon; other study pressures experienced by participants but unknown to the research team may have contributed to participants' overall stress levels at the completion of the course. Some members of the group did not regularly attend the mindfulness training sessions and a few had dropped out by the end of the eight week course. It was not possible to monitor compliance with the homework exercises, and the regularity and quantity of home practice is unknown. Lack of these data may contribute to uneven results. Furthermore, respiration patterns were not screened for at the start of the study. Unstable breathing patterns (such as a tendency to hyperventilate) are known to cause fluctuations in heart rate variability (McCraty et al., 2009). Ideally, HRV coherence recordings would have been measured in a quiet space and a stable baseline ensured before recordings began. No medical history was taken from participants and any other possible physiological influences are unknown.

However, not all extraneous variables would necessarily affect the results in a negative direction. Participants were not screened for participation in other stress-reduction practices, such as other forms of meditation, yoga, tai chi etc. It is known that one participant practises daily biofeedback, several members of the group were regular meditators prior to the mindfulness course, while another intimated being an experienced practitioner of yoga. It should also be noted that while the findings of this study ran in the "wrong" direction on both measured variables, the extent to which they did so was very small.

While certain aspects of the methodology of this study may have contributed to the anomalous findings, the main reason the results should be interpreted with caution is because of the very small sample size ($N$=13). The study therefore lacked statistical power and, in any case, despite the unexpected outcome the results were not significant. Consequently, it is not possible to draw any practical or theoretical conclusions from the findings of this study.

*Case Study*

The participant selected for this case study, Participant No. 10, a 24 year old female, had PSS scores and HRV coherence scores that were fairly typical for the sample. Her PSS score before training was a little higher than average for the group, decreasing very slightly after training. This participant's PSS score after training was very close to the mean for the group, but the group mean had risen. Participant No. 10's PSS scores were within the published norms for a respondent of her age both before and after the training suggesting that her levels of perceived stress were no higher nor lower than to be expected.

Participant No. 10's HRV coherence scores were in a low range, as were those for all the participants. However, low coherence measurements on the emWave2 are not unusual in people who have never undergone biofeedback training, and coherence generally rises with practice.  This study did not incorporate any biofeedback training, and the emWave2 was used purely as a measurement tool.  Participant No. 10's coherence score decreased somewhat after training, which would usually suggest less harmonious ANS activity, and therefore higher stress levels (McCraty et al., 2009).  However, this participant's mean heart rate reduced considerably after the training, and the mean R-R interval was more varied.  These latter indicators suggest a healthier, less stressed heartbeat rhythm.

While these findings may seem contradictory, as mentioned previously, coherence measured by the emWave2 actually measures three different factors – heart rhythms, breathing patterns, and blood pressure rhythms – and mathematically combines them into one overall coherence score.  The most likely explanation for the discrepancy between the R-R variance and HRV coherence score is a difference in breathing pattern, since breathing directly influences coherence scores.  Indeed, learning breath control is fundamental to HRV biofeedback training and learning to control ANS activity (McCraty et al., 2009).  Participants were not screened for unstable breathing patterns and this may be a point to address in future research.

Muscle artefact can also artificially depress coherence scores (McCraty et al., 2009), and while recordings made using the emWave Pro desktop software with a dongle plugged directly into a PC deartefacts what it detects as obvious muscle movements, the handheld emWave2 does not possess this capability.  The extent to which artefact may have impacted on Participant No. 10's HRV coherence measurements cannot be determined from inspecting the data.  A future study may wish to use the emWave Pro or emWave Pro Plus PC desktop versions to measure HRV coherence to reduce muscle movement artefact in the data.

### Conclusions and Future Research

This study contradicted the findings of earlier literature inasmuch as perceived stress increased and HRV coherence decreased after the eight week Mindfulness-Based Cognitive Therapy course.  However, the study lacked statistical power and the results were not statistically significant.  No definitive conclusions can therefore be drawn on the effects of mindfulness training on stress and heart rate variability based on the current study.

Future research should address the methodological limitations outlined above and use a larger sample to overcome the lack of statistical power.  Further investigation into the physiological correlates of mindfulness training may also be a worthwhile line of enquiry, either using other measures of ANS activity such as galvanic skin response, or a measure of CNS activity such as EEG, with a view to investigating the psychophysiological mechanisms of mindfulness.  Research should focus on how such an understanding may inform more effective applications of mindfulness, which specific techniques are most beneficial and for which populations.

## APPENDICES

**Appendix A:** Participant Information Sheets
**Appendix B:** Participant Consent Form
**Appendix C:** Perceived Stress Scale Questionnaire
**Appendix D:** SPSS Output

# APPENDIX A – PARTICIPANT INFORMATION SHEETS

**EXAMINING THE EFFECTS OF MINDFULNESS ON LEVELS OF PERCEIVED STRESS**

We are carrying out a study on levels of perceived stress as part of the Clinical Applications of Mindfulness module.

For our study, we would like to ask you to complete a questionnaire. The questionnaire is only 10 questions long. Please read the instructions carefully at the top of the questionnaire before answering. You will also be asked some basic information about you, such as your age and your gender.

The information we collect will be uploaded to a database that will be shared among our research project team, but the data you provide will be kept completely anonymously. This means that there will be no reference to your name or any other personal details that would enable anyone to identify you.

Please note that your participation in this study is voluntary. If you would like to end the experiment at any point you are free to do so, and you do not need to offer an explanation. You may also withdraw your participation after the questionnaires have been completed and collected in, up until the point where your data is combined with that of other participants, as after that it will not be possible to determine which data is yours. If you do choose to withdraw from the study before that point, any data collected from you will be removed from the dataset.

Please do not hesitate to ask if you have any questions or if anything you have been asked to do is not clear.

Please read the consent form carefully and sign it. This is to confirm that you have understood the information provided above, and that you are aware that you have the right to withdraw from the experiment at any point.

Thank you for your participation and we hope you enjoy the mindfulness sessions.

**Levels of Perceived Stress Research Team**

**INFORMATION ABOUT THE EMWAVE**

This is an additional part of the research our group is carrying out as part of the effects of mindfulness on levels of perceived stress project, which is being conducted as my own personal part of the study.

I will be using a device called the emWave®, which uses an infrared ear clip that picks up your pulse and the machine detects the patterns of your heartbeat and electrical rhythms of the heart. There is no physical sensation while using the machine and you will not be required to do anything except sit still while the data recording takes place. I will be recording two minutes of data.

This information will be uploaded to a PC with software for analysing the recorded data. This information will be kept by me and will not be shared with anyone else. Once your data is recorded there is no way of identifying you from the data.

As with participation in the questionnaire, your participation in this part of the research is voluntary.

I am very happy to answer any questions you may have about the emWave device, its use in this study, and its wider use as a stress reduction and performance coaching tool.

**Gwyneth Rolph**

# APPENDIX B – CONSENT FORM

**EXAMINING THE EFFECTS OF MINDFULNESS ON LEVELS OF PERCEIVED STRESS**

I have read the Participation Information Sheet and I am willing to participate in the research study.

Print name ________________________________________

Signature ________________________________________ Date __________

All reasonable steps have been taken to provide an appropriate explanation of the research to the participant

Signed ____________________________ Researcher__________________ Date _______

## PERCEIVED STRESS SCALE

The questions in this scale ask you about your feelings and thoughts during the last month. In each case, you will be asked to indicate by circling *how often* you felt or thought a certain way.

Name _________________________________________________ Date ______________

Age __________ Gender (Circle): **M  F** Other _________________________________

### 0 = Never 1 = Almost Never 2 = Sometimes 3 = Fairly Often 4 = Very Often

1.  In the last month, how often have you been upset because of something that happened unexpectedly? ................................................. 0   1   2   3   4

2.  In the last month, how often have you felt that you were unable to control the important things in your life? ................................................. 0   1   2   3   4

3.  In the last month, how often have you felt nervous and "stressed"? ....... 0   1   2   3   4

4.  In the last month, how often have you felt confident about your ability to handle your personal problems? ................................................. 0   1   2   3   4

5.  In the last month, how often have you felt that things were going your way? ................................................................................... 0   1   2   3   4

6.  In the last month, how often have you found that you could not cope with all the things that you had to do? ..................................................... 0   1   2   3   4

7.  In the last month, how often have you been able to control irritations in your life? ........................................................................... 0   1   2   3   4

8.  In the last month, how often have you felt that you were on top of things? ................................................................................... 0   1   2   3   4

9.  In the last month, how often have you been angered because of things that were outside of your control? ................................................. 0   1   2   3   4

10. In the last month, how often have you felt difficulties were piling up so high that you could not overcome them? ........................................... 0   1   2   3   4

info@mindgarden.com
www.mindgarden.com

# PERCEIVED STRESS SCALE
## by Sheldon Cohen

The Perceived Stress Scale (PSS) is the most widely used psychological instrument for measuring the perception of stress. It is a measure of the degree to which situations in one's life are appraised as stressful. Items were designed to tap how unpredictable, uncontrollable, and overloaded respondents find their lives. The scale also includes a number of direct queries about current levels of experienced stress. The PSS was designed for use in community samples with at least a junior high school education. The items are easy to understand, and the response alternatives are simple to grasp. Moreover, the questions are of a general nature and hence are relatively free of content specific to any subpopulation group. The questions in the PSS ask about feelings and thoughts during the last month. In each case, respondents are asked how often they felt a certain way.

**Evidence for Validity**: Higher PSS scores were associated with (for example):
- failure to quit smoking
- failure among diabetics to control blood sugar levels
- greater vulnerability to stressful life-event-elicited depressive symptoms
- more colds

**Health status relationship to PSS**: Cohen et al. (1988) show correlations with PSS and: Stress Measures, Self-Reported Health and Health Services Measures, Health Behavior Measures, Smoking Status, Help Seeking Behavior.

**Temporal Nature**: Because levels of appraised stress should be influenced by daily hassles, major events, and changes in coping resources, predictive validity of the PSS is expected to fall off rapidly after four to eight weeks.

**Scoring**: PSS scores are obtained by reversing responses (e.g., 0=4, 1=3, 2=2, 3=1 & 4=0) to the four positively stated items (items 4, 5, 7, & 8) and then summing across all scale items. A short 4 item scale can be made from questions 2, 4, 5 and 10 of the PSS 10 item scale.

**Norm Groups**: L. Harris Poll gathered information on 2,387 respondents in the U.S.

### Norm Table for the PSS 10 item inventory

| Category | N | Mean | S.D. |
|---|---|---|---|
| **Gender** | | | |
| Male | 926 | 12.1 | 5.9 |
| Female | 1406 | 13.7 | 6.6 |
| **Age** | | | |
| 18-29 | 645 | 14.2 | 6.2 |
| 30-44 | 750 | 13.0 | 6.2 |
| 45-54 | 285 | 12.6 | 6.1 |
| 55-64 | 282 | 11.9 | 6.9 |
| 65 & older | 296 | 12.0 | 6.3 |
| **Race** | | | |
| White | 1924 | 12.8 | 6.2 |
| Hispanic | 98 | 14.0 | 6.9 |
| Black | 176 | 14.7 | 7.2 |
| other minority | 50 | 14.1 | 5.0 |

# APPENDIX D: SPSS Output

**Table 3: Perceived Stress Scale Questionnaire Scores – Before and After Mindfulness Training**

**Descriptives**

| | | | Statistic | Std. Error |
|---|---|---|---|---|
| Perceived Stress Scale Score Before Training | Mean | | 15.7778 | 1.96340 |
| | 95% Confidence Interval for Mean | Lower Bound | 11.2502 | |
| | | Upper Bound | 20.3054 | |
| | 5% Trimmed Mean | | 16.0309 | |
| | Median | | 17.0000 | |
| | Variance | | 34.694 | |
| | Std. Deviation | | 5.89020 | |
| | Minimum | | 4.00 | |
| | Maximum | | 23.00 | |
| | Range | | 19.00 | |
| | Interquartile Range | | 7.50 | |
| | Skewness | | -1.102 | .717 |
| | Kurtosis | | .963 | 1.400 |
| Perceived Stress Scale Score After Training | Mean | | 16.1111 | 2.94602 |
| | 95% Confidence Interval for Mean | Lower Bound | 9.3176 | |
| | | Upper Bound | 22.9046 | |
| | 5% Trimmed Mean | | 16.0679 | |
| | Median | | 17.0000 | |
| | Variance | | 78.111 | |
| | Std. Deviation | | 8.83805 | |
| | Minimum | | .00 | |
| | Maximum | | 33.00 | |
| | Range | | 33.00 | |
| | Interquartile Range | | 8.00 | |
| | Skewness | | .082 | .717 |
| | Kurtosis | | 2.084 | 1.400 |

**Table 4: Paired t-test – Perceived Stress Scale questionnaire scores before and after mindfulness training**

**Paired Samples Test**

| | | Paired Differences | | | | | t | df | Sig. (2-tailed) |
|---|---|---|---|---|---|---|---|---|---|
| | | | | | 95% Confidence Interval of the Difference | | | | |
| | | Mean | Std. Deviation | Std. Error Mean | Lower | Upper | | | |
| Pair 1 | Perceived Stress Scale Score Before Training - Perceived Stress Scale Score After Training | -.33333 | 5.80948 | 1.93649 | -4.79889 | 4.13222 | -.172 | 8 | .868 |

**Table 5: Heart Rate Variability coherence scores before and after mindfulness training**

**Descriptives**

| | | | Statistic | Std. Error |
|---|---|---|---|---|
| Heart Rate Variability Before Training | Mean | | 1.0125 | .25102 |
| | 95% Confidence Interval for Mean | Lower Bound | .4189 | |
| | | Upper Bound | 1.6061 | |
| | 5% Trimmed Mean | | .9694 | |
| | Median | | .9000 | |
| | Variance | | .504 | |
| | Std. Deviation | | .71001 | |
| | Minimum | | .30 | |
| | Maximum | | 2.50 | |
| | Range | | 2.20 | |
| | Interquartile Range | | .93 | |
| | Skewness | | 1.377 | .752 |
| | Kurtosis | | 2.447 | 1.481 |
| Heart Rate Variability After Training | Mean | | .9500 | .31396 |
| | 95% Confidence Interval for Mean | Lower Bound | .2076 | |
| | | Upper Bound | 1.6924 | |
| | 5% Trimmed Mean | | .8778 | |
| | Median | | .6500 | |
| | Variance | | .789 | |
| | Std. Deviation | | .88802 | |
| | Minimum | | .20 | |
| | Maximum | | 3.00 | |
| | Range | | 2.80 | |
| | Interquartile Range | | .72 | |
| | Skewness | | 2.147 | .752 |
| | Kurtosis | | 5.098 | 1.481 |

**Table 6: Wilcoxon Signed Ranks Test – HRV Coherence scores before and after mindfulness training**

**Ranks**

| | | N | Mean Rank | Sum of Ranks |
|---|---|---|---|---|
| Heart Rate Variability After Training - Heart Rate Variability Before Training | Negative Ranks | 5[a] | 4.00 | 20.00 |
| | Positive Ranks | 3[b] | 5.33 | 16.00 |
| | Ties | 0[c] | | |
| | Total | 8 | | |

a. Heart Rate Variability After Training < Heart Rate Variability Before Training

b. Heart Rate Variability After Training > Heart Rate Variability Before Training

c. Heart Rate Variability After Training = Heart Rate Variability Before Training

**Test Statistics[a]**

| | Heart Rate Variability After Training - Heart Rate Variability Before Training |
|---|---|
| Z | -.280[b] |
| Asymp. Sig. (2-tailed) | .779 |

a. Wilcoxon Signed Ranks Test

b. Based on positive ranks.